\documentclass{emulateapj}
\shorttitle{A New Classification Method For GRBs}
\shortauthors{L\"u et al.}
\slugcomment{}
\begin{document}

\title{A New Classification Method for Gamma-Ray Bursts}
\author{Hou-Jun L\"{u}\altaffilmark{1}, En-Wei Liang\altaffilmark{1,2}, Bin-Bin
Zhang\altaffilmark{2}, and Bing Zhang\altaffilmark{2}
}\altaffiltext{1}{Department of Physics, Guangxi University, Nanning 530004,
China; lew@gxu.edu.cn}\altaffiltext{2}{Department of Physics and Astronomy,
University of Nevada, Las Vegas, NV 89154. zhang@physics.unlv.edu}

\begin{abstract}
Recent {\em Swift} observations suggest that the traditional long vs. short GRB
classification scheme does not always associate GRBs to the two physically
motivated model types, i.e. Type II (massive star origin) vs. Type I (compact
star origin). We propose a new phenomenological
classification method of GRBs by introducing a
new parameter $\varepsilon=E_{\gamma, \rm iso,52}/E^{5/3}_{p,z,2}$, where
$E_{\gamma,\rm iso}$ is the isotropic gamma-ray energy (in units of
$10^{52}$erg), and $E_{\rm p,z}$ is the cosmic rest frame spectral peak energy
(in units of 100 keV). For those short GRBs with ``extended emission", both
quantities are defined for the short/hard spike only. With the current complete
sample of GRBs with redshift and $E_p$ measurements, the $\varepsilon$
parameter shows a clear bimodal distribution with a separation at $\varepsilon
\sim 0.03$. The high-$\varepsilon$ region encloses the typical long GRBs with
high-luminosity, some high-$z$ ``rest-frame-short" GRBs (such as GRB 090423 and
GRB 080913), as well as some high-$z$ short GRBs (such as GRB 090426). All
these GRBs have been claimed to be of the Type II origin based on other
observational properties in the literature. All the GRBs that are argued to be
of the Type I origin are found to be clustered in the low-$\varepsilon$ region.
They can be separated from some nearby low-luminosity long GRBs (in $3\sigma$)
by an additional $T_{90}$ criterion, i.e. $T_{90,z}\lesssim 5$ s in the {\em
Swift}/BAT band. We suggest that this new classification scheme can better
match the physically-motivated Type II/I classification scheme.
\end{abstract}


\section{Introduction}
Phenomenologically, gamma-ray bursts (GRBs) are classified as long {\em vs.}
short with a division line at the observed duration $T_{90}\sim 2$ s
(Kouveliotou et al. 1993). Robust associations of the underlying supernovae
(SNe) with some long GRBs (Galama et al. 1998; Stanek et al. 2003; Hjorth et
al. 2003; Malesani et al. 2004; Modjaz et al. 2006; Pian et al. 2006) and the
fact that long GRB host galaxies are typically irregular galaxies with intense
star formation (Fruchter et al. 2006) suggest that they are likely related to
the deaths of massive stars, and the $``$collapsar$"$ model has been widely
recognized as the standard scenario for long GRBs (Woosley 1993; Paczy\'{n}ski
1998; Woosley \& Bloom 2006). Observational breakthroughs led by the {\em
Swift} mission (Gehrels et al. 2004) suggest that at least some short GRBs are
associated with nearby host galaxies with little star formation (Tanvir et al.
2005), and that they are not associated with an underlying SN (Gehrels et al.
2005; Villasenor et al. 2005; Fox et al. 2005; Hjorth et al. 2005; Berger et al. 2005), favoring
the idea that they are produced by mergers of two compact stellar objects, such
as NS-NS and NS-BH mergers (Eichler et al. 1989; Narayan et al. 1992; Nakar
2007).

However, several lines of observational evidence in the {\em Swift} era
suggest that duration is not necessarily a reliable indicator of the physical
nature of a GRB. (1) The non-detection of a SN signature associated with the
nearby long GRBs 060614 and 060505 (Gehrels et al. 2006; Gal-Yam et al. 2006;
Fynbo et al. 2006; Della Valle et al. 2006) disfavors the conventional
collapsar scenario of long GRBs (Woosley \& Bloom 2006). Some observational
properties of the $\sim100$ s long GRB 060614 are very similar to those of some
short GRBs (Gehrels et al. 2006; Zhang et al. 2007), making it likely
associated with the physical category that most short GRBs belong to. (2)
Significant soft ``extended" gamma-ray emission and late X-ray flares are
observed in a handful of ``short" GRBs (Barthelmy et al. 2005; Norris \&
Bonnell 2006; Perley et al. 2009), suggesting that they are not necessarily
short. A new physically motivated classification scheme,
i.e., Type II (massive star origin) vs. Type I (compact star origin) was
proposed (Zhang 2006; Zhang et al. 2007). Kann et al (2010, 2008) systematically
studied the optical afterglow emission properties of Type II and Type I GRBs,
and suggested that these properties carry important information about the nature
of GRB progenitors. A full definition of Type I/II GRBs as well as the multiple
observational criteria are presented in Zhang et al. (2009). (3)
Two high-redshift gamma-ray bursts, GRB 080913 at $z=6.7$ (Greiner et al.
2009; Perez-Ramirez et al. 2010) and GRB 090423 at $z=8.2$ (Tanvir et al. 2009;
Salvaterra et al. 2009), appear as intrinsically short GRBs. Their observed durations
are $(8\pm 1)$s and $\sim10.3$s in the {\em Swift/BAT} band, respectively. Corrected
redshift, the durations of two high-$z$ GRBs are shorter than 2 seconds in the rest
frame. The physical origin of these high-$z$ GRBs have been subject to debate
(Greiner et al. 2009; Perez-Ramirez er al. 2010; Tanvir et al. 2009; Salvaterra et
al.2009). Although compact star mergers may occur at such a high redshift
(Belczynski et al. 2010), various observational properties of these two GRBs
point towards the massive star origin (Zhang et al. 2009; Lin et al. 2009;
Belczynki et al. 2010). (4) A more striking case is GRB 090426, whose observed BAT
band $T_{90}$ is only $1.2\pm 0.3$s, and the rest frame duration is only $\sim$ 0.33 s
at $z=2.609$ (Levesque et al. 2009). It greatly exceeds the previous short GRB
redshift record, i.e. $z = 0.923$ for GRB 070714B (Graham et al. 2009).
Phenomenologically, this is an unambiguous short-duration GRB, but the available
afterglow and host galaxy properties point towards a different picture:
it has a blue, very luminous, star-forming putative
host galaxy with a small angular offset of the afterglow location from the
center, and a similar medium density as typical Type II GRBs. All these
suggest that the burst is more closely related to the core collapse of a massive
star (Levesque et al. 2009; Antonelli et al. 2009; Xin et al. 2010). (5) Monte Carlo
simulations suggest that the compact star merger model cannot interpret both
the {\em Swift} $z$-known short GRB sample and the BATSE short GRB sample. Instead
one may need a mix of GRBs from compact star
mergers and from massive star core collapses to account for the observed
short GRB population (Virgili et al. 2009; Cui et al.
2010).

In summary, the traditional long {\em vs.} short GRB scheme no longer always
correspondingly match the two distinct physical origins, i.e., collapse of massive
stars (Type II GRBs) {\em vs.} mergers of compact stars (Type I GRBs) (Zhang 2006;
Zhang et al. 2007, 2009; Bloom et al. 2008). Zhang et al (2009) proposed a
procedure (a flow chart, see their Fig.8) invoking a full set of observational
characteristics, including supernova(SN)-GRB association, specific star-forming
rate (SFR) of the host galaxy, burst location offset, burst duration, hardness,
spectral lag, statistical correlations, energetics, and afterglow properties,
to judge the physical category of a GRB.
On the other hand, for most GRBs one may not immediately retrieve all the
information needed for such a classification. It would be interesting to search
for additional phenomenological classification schemes to see whether there exist
other quantities, especially those invoking prompt GRB emission properties only,
that can give a good indication of the physical nature of a GRB.
This motivates us to explore a new phenomenological classification method.
Besides the burst duration, we think both the burst energy and the spectral
properties in the burst rest frame would carry important information about
the nature of the burst. In this paper, we propose a new discriminator based on
the isotropic burst energy and rest frame peak energy ($E_{\rm p}$) of the $\nu f_\nu$
spectrum of the prompt gamma-rays. We will show that this classification method
is more closely connected to the Type II/I physical classification scheme.

\section{Sample}
In order to develop our classification method, we use a sample of all GRBs with
both redshift and spectral parameters known up to Mar, 2010. Our sample
includes 137 GRBs. They are detected by {\em BeppoSAX}, {\em HETE}-2, {\em
Swift}, {\em Suzaku}, and {\em Fermi}. Most of them are taken from Amati et al.
(2008, 2009), Kann et al. (2010), Krimm et al. (2009), Zhang et al. (2009) and
references therein. Their isotropic gamma-ray energy ($E_{\gamma,iso}$) is
calculated in the GRB rest-frame $1-10^4$ kev band with the spectral parameters
in order to avoid instrument selection effect. The soft XRF 080109 (Soderberg et
al. 2008), whose emission is in the XRT band (0.3-10kev) instead of BAT band,
is also included in our sample. The XRT and UVOT observations place a limit of
its $E_p$ in the range $0.037~{\rm kev} < E_{peak} < 0.3~{\rm kev}$. We adopt
$E_{peak}\approx0.12^{+0.23}_{-0.089}$ kev (Li 2008). Eight bursts in our
sample (GRBs 050709, 050724, 051210, 060614, 061006, 061210, 070714B, and
071227) are the so-called short GRBs with ``extended emission" discussed in
literature. For these GRBs, both parameters ($E_{\gamma,iso}$ and $E_{p,z}$)
are derived for the initial hard spike only (extended emission
excluded)\footnote{We note that $T_{90}$ depends on the detector's sensitivity
and energy band. In our analysis, we define $T_{90}$ using the BAT energy band
and sensitivity. In the {\em Swift} era, it is found that some ``short" GRBs
(as observed in the BATSE band) have extended emission. It is the convention
of the {\em Swift} team and the community to define a ``short" GRB based on
the duration of the short spike only, with the extended emission excluded.
Observationally, the spectrum of the short spike is much harder than that of
the extended emission, the latter shares many properties with X-ray flares.
For a same burst, the ``extended emission" depends on redshift, since it may
be buried within the background if the redshift is high enough. We therefore
only use the information of the short/hard spike to perform our classification.}.
We tabulate only those GRBs with $T_{90,z}=T_{90}/(1+z) < 2$ s (Twenty-nine GRBs)
in our sample in Table 1.

\section{A New GRB Classification Parameter}
Firstly, we show the distribution of $T_{90,z}$ in Fig. 1(a). Similar to that
shown in Lin et al. (2009), $T_{90,z}$ for the current GRB sample with redshift
measurements has a tentative bimodal distribution, with peaks at $\log
T_{90,z}=-0.5$ and 1.25, respectively. Due to the overlap between the two
classes (long vs. short), those GRBs having $T_{90,z}=1\sim 5$ seconds cannot
be unambiguously categorized into either group\footnote{It was proposed that
those GRBs having $T_{90,z}=1\sim 5$ seconds may be an intermediate population
based on the excess of the GRBs over the bimodal distribution of $T_{90,z}$.
Although these GRBs are less energetic and have dimmer afterglows than typical
long GRBs, they share many similar properties of typical GRBs, such as spectral
lag and spectrum-energy correlations. They may be a sub-group of the long GRBs
(de Ugarte Postigo 2010).}. We test the bimodality with the KMM algorithm
presented by Ashman et al. (1994), and obtain a chance probability $p \sim
10^{-2}$. The correct allocation rate for short GRBs estimated with the KMM
algorithm is $73.3\%$.

We develop our classification method by defining a parameter,
\begin{equation}
\varepsilon (\kappa) \equiv E_{\gamma,\rm iso,52}/E_{\rm p,z,2}^\kappa,
\end{equation}
where $E_{\gamma,\rm iso, 52}=E_{\gamma,\rm iso}/10^{52}$ ergs and $E_{\rm
p,z}=E_p (1+z)/10^2$ keV. Note that $\varepsilon$ does not depend on any
spectrum - energy correlation or theoretical assumption. We explore various
$\kappa$ values, and found that given a same observed GRB (known fluence and
$E_p$) the implicit $z$-dependence of $\varepsilon(\kappa) \propto
D^2_L(z)/(1+z)^{1+\kappa}$ essentially vanishes at $z>2$, if $\kappa=5/3$ (see
Fig. 2 for a comparison of the $z$-dependence of $\varepsilon$ for different
$\kappa$). We therefore define
\begin{equation}
\varepsilon \equiv E_{\gamma,\rm iso,52}/E_{\rm p,z,2}^{5/3},
\end{equation}
which removes the redshift dependence for high-$z$ GRBs. This parameter may be
more suitable for the classification of high-$z$ GRBs, such as GRBs 090423 and
080913.

We calculate $\varepsilon$ for the bursts in our sample and show the $\log
\varepsilon$ distribution in Fig.1 (b). A clear bimodal feature is found, with a
division line at $\sim 0.03$ (high-$\varepsilon$ vs. low-$\varepsilon$). The
high-$\varepsilon$ portion follows a log normal distribution with
$\varepsilon=0.80\pm 0.11$ ($1\sigma$). For the low-$\varepsilon$ it is
$0.003\pm 0.002$. We test the bimodality of the $\varepsilon$ distribution with
the KMM algorithm and obtain a chance probability $p<10^{-4}$. The overall
correct allocation rate of GRBs in the two categories (high-$\varepsilon$ vs.
low-$\varepsilon$) is as high as 99.8\%, indicating that $\varepsilon$ is
a new parameter for GRB classification.

It is generally believed that GRBs originate either from core collapses of
massive stars (Type II) or mergers of compact stars (Type I). One may ask
whether this phenomenal classification matches the physical scheme of the Type
I and Type II. The origins of most GRBs in our sample have been extensively
discussed in literature (Zhang et al. 2009; Kann et al. 2010, 2008). We
therefore examine how the members in our two new categories (high-$\varepsilon$
vs. low-$\varepsilon$) are associated with the two physical model types. To
define the physical category of the GRBs, we apply the flow chart of Zhang et
al. (2009, their Fig.8). We denote all ``Type II" or ``Type II candidates" as
``Type II", and ``Type I" or ``Type I candidates" as ``Type I" (which include
the Type I Gold sample and most other short/hard GRBs in Zhang et al. 2009). We
note that these definitions are fully consistent with those adopted in Kann et
al. (2010, 2008). Fig. 1 displays all the GRBs in our sample in the
$\varepsilon-T_{90,z}$ plane. Different symbols are used to mark different
groups. For example, triangles denote the GRBs that are believed to be of the
Type II origin, among which the black triangles denote the traditional long
GRBs, and the blue triangles denote the GRBs those with $T_{90,z}$'s shorter
than 2 s. The red circles denote those GRBs that are believed to have the Type
I origin. Six green stars denote nearby low luminosity GRBs with supernova
associations. It is remarkable to see that the $\varepsilon$ classification
clearly separates Type II GRBs from Type I GRBs.

In order to quantitatively assess the clustering feature among different types
of GRBs, we derive the GRB distribution probability $p (\log \varepsilon, \log
T_{90,z})$ from the ratio between the number of GRBs in the grid $\log
\varepsilon+d\log \varepsilon , \log T_{90,z}+d\log T_{90,z}$ and the total
number of GRBs in our sample. Since the distribution of $\varepsilon$ is
concerned in our analysis, we take the error of $\log \varepsilon$ into account
in our calculation of $p$. For a given GRB with $\log \varepsilon\pm \delta\log
\varepsilon$, we assume that the probability distribution of this GRB is a
normalized log-normal distribution centering at $\log \varepsilon$ with a width
of $\delta\log \varepsilon$. The probability ($p_i$) of the GRB in a given rage
[$\log \varepsilon-d\log \varepsilon$/2, $\log \varepsilon+d \log
\varepsilon$/2] is calculated by integrating the probability distribution
function in the range. The probability $p (\log \varepsilon, \log T_{90,z})$ is
then obtained by summing up $p_i$ over the GRBs in our sample. We take $d\log
T_{90,z}=0.35$ and $d \log \varepsilon=0.60$ in our calculation. The contours
of the probability distribution are shown in Fig. 1(c). One can clearly
identify two clustering regions [with $p (\log \varepsilon, \log
T_{90,z})>0.1$] centered in the high-$\varepsilon$ regime and the
low-$\varepsilon$, low-$T_{90,z}$ regime. At the $3\sigma$ level, the
high-$\varepsilon$ group includes typical long GRBs with high-luminosity, some
intrinsic short duration GRBs (including the high-$z$ GRB 090423 and GRB
080913), as well as some high-$z$ short GRBs (e.g. GRB 090426). In contrast,
the low-$\varepsilon$ group contains the extensively discussed Type I GRBs,
including those with and without extended emission. Based on the probability
contours, a low-$\varepsilon$ GRB with $T_{90,z}\lesssim 5$ s in the {\em
Swift}/BAT band would be identified as a Type I GRB at the $3\sigma$ confidence
level. GRBs 060614 and 060505 are marginally included in the $p (\log
\varepsilon, \log T_{90,z})> 0.003$ region (the light grey region in Fig. 1c)
of the Type I GRBs. Nearby low luminosity long GRBs (LL-GRBs), e.g. GRBs
980425, 031203, 050826, and 060218 (green stars in Fig. 1c) are out of the 3
$\sigma$ contours of the Type I and Type II GRBs. This may hint a distinct GRB
population as suggested by Liang et al. (2007; see also Soderberg et al. 2004;
Cobb et al. 2006).

\section{Conclusions and Discussion}
By introducing a new parameter $\varepsilon$, we have proposed a new
phenomenological classification method for GRBs. We demonstrated that GRBs with
known redshifts are cozily classified into two classes with a separation
$\varepsilon=0.03$. Due to the clear bimodal distribution of $\varepsilon$ with
little overlap, the overall correct allocation rate of a GRB to a particular
category (high-$\varepsilon$ vs. low-$\varepsilon$) is as high as $>99.8\%$,
which is much better than the traditional duration classification method. More
importantly, this classification scheme is more closely related to the two
physically motivated model classes - Type II vs. I. In particular, the
high-$\varepsilon$ category is a good representation of the Type II GRBs, while
the low-$\varepsilon$ category, with an additional duration criterion
$T_{90,z}\lesssim 5$ s, is a good representation of the Type I GRBs. We suggest
that the $\varepsilon$ parameter can be evaluated for future GRBs with $z$
measurements, and the $\varepsilon$-based classification may be performed
regularly to infer the physical origin of a GRB.

The well-separated bimodal distribution of $\log \varepsilon$ depends on our
selection of $\kappa$ in Eq. (1). As shown in Fig. 2, $\varepsilon$ is
insensitive to $z$ at $z>2$ for a given burst by taking $\kappa=5/3$, but it is
not at low redshift. One may suspect that if the bimodal distribution is due to
the redshift effect. To examine this issue, we re-plot the Figure 1 by separating
GRBs into the $z>2$ and the $z<2$ samples (Figure 3, left and middle panels,
respectively). We also show the GRB distributions
in the $\log (1+z)-\log (\varepsilon)$ plane for all the bursts in our sample
in Figure 3. It is found that the the clustering feature of $\log \varepsilon$ is
not caused by the redshift selection effect. Therefore, $\varepsilon$ is a GRB
discriminator for both GRBs at $z<2$ and $z>2$.

Another concern about our classification method may be sample uniformity.
The GRBs in our sample were detected with different instruments with different
instrumental threshold. However, our
classification method is based on the global energy and spectral information.
Different from the burst duration, a well-measured $E_p$ essentially does not
depend on the detector. So is $E_{\rm \gamma,iso}$. The $E_p$ values of the
GRBs in our sample, which expand almost four orders of magnitude, are well
within the energy band of previous and current GRB missions, so they are well
measured. The $E_{\rm iso}$ values expand almost eight
orders of magnitude. Therefore, our sample could be regarded as a complete
sample with known $E_p$ and $E_{\rm iso}$ for current missions.

We should note that our classification scheme is particularly helpful to
diagnose the physical origin of GRBs with a rest-frame short duration (e.g.
$T_{90,z} < 2$ s, Table 1). For example, it convincingly classifies GRBs
090423, 080913 and 090426 into the high-$\varepsilon$, and therefore, the Type
II category, which is otherwise difficult with the duration
criterion\footnote{As shown in Figure 2, a GRB with the same observed fluence
and $E_p$ at higher redshift may have a higher $\varepsilon$. However, the
bimodal distribution may not be due to the redshift effect (see Figure 3).}. The
distinction of the $T_{90,z} < 2 s$ bursts with different $\varepsilon$ values is
also evident from their X-ray afterglow properties. Kann et al. (2008) compared
the optical afterglows Type I GRBs with Type II GRBs, and found that those of
Type I GRBs have a lower average luminosity and show a larger intrinsic spread
of luminosities. In Figure 4, we compare the XRT lightcurves and the 12-hour
X-ray luminosities between the two types of $T_{90,z}<5 s$ s GRBs. The
low-$\varepsilon$ $T_{90,z}<2 s$ s GRBs (Type I, red) are systematically fainter
than the high-$\varepsilon$ ones (Type II, blue)\footnote{We cannot exclude the
possibility that this is a selection effect, since high-$z$ Type I GRBs may have
fainter X-ray afterglows and therefore may
not be detected.}, whose properties are
rather similar to other high-$\varepsilon$ long GRBs (black).

Physically it is not obvious why the parameter $\varepsilon$ (corresponding to
$\kappa=5/3$) gives a cozy classification scheme. We have tried other $\kappa$
values in Eq.(1), but the classifications defined by other
$\varepsilon(\kappa)$ are not as clean as the one defined by Eq.(2). As shown
in Figure 2, an apparent advantage of $\kappa=5/3$ is to diminish the
$z$-dependence at high-$z$ for a given observed GRB.

One can also discuss the connection between the log-normal distribution of
$\varepsilon$ in the high-$\varepsilon$ regime and some empirical relations of
Type II GRBs. With the Amati-relation $E_{p,z}\propto E_{\gamma,\rm
iso}^{0.54}$ (Amati et al. 2009), we find that $\varepsilon\propto
E^{0.185}_{\rm p, z}$, insensitive to $E_{\rm p,z}$. Liang \& Zhang (2005)
discovered an empirical relation among $E_{\gamma,\rm iso}$, $E_{\rm p, z}$,
and $t_{\rm b,z}$, namely, $E_{\gamma,\rm iso}\propto E^{1.94 \pm 0.17}_{\rm p,
z}t^{-1.24 \pm 0.23}_{\rm b,z}$, where $t_{\rm b, z}$ is the rest-frame break
time of the optical afterglow lightcurve. This gives $\varepsilon\propto E_{\rm
p,z}^{0.27\pm 0.17} t^{-1.24\pm 0.23}_{\rm b, z}$, which is sensitive to
$t_{\rm b,z}$. A clustering in $\varepsilon$ therefore corresponds to a
clustering in $t_{\rm b,z}$. Interpreting the $t_{\rm b,z}$ as the jet break
time, the Liang-Zhang relation may be translated into the Ghirlanda-relation,
$E_{\gamma,{\rm iso},52} (1-cos \theta_j)\ \simeq AE_{\rm p,z,2}^{\sim 1.7}$,
where $A \sim 0.01-0.03$ (Ghirlanda et al. 2004; Dai et al. 2004). Given
$\varepsilon \equiv E_{\gamma,{\rm iso},52}/E_{\rm p,z,2}^{1.7}$,  we can get
$\varepsilon\approx A (1-\cos\theta_j)^{-1}$, where $\theta_j$ is the jet
opening angle. The lognormal $\varepsilon$ distribution ($\varepsilon =0.80\pm
0.11$, at $1\sigma$) therefore corresponds to a lognormal jet opening angle
distribution for Type II GRBs, i.e., $\theta_j\sim 0.16\pm 0.03$ rad (assuming
$A\sim 0.01$). We note that the jet opening angle derived by Ghirlanda et al.
(2004) indeed clusters in a small range. Since the Ghirlanda-relation is very
difficult to understand physically, we suspect that it may be related the fact
that $\varepsilon$ is quasi-universal for Type II GRBs. Future data with very
early or very late optical break times may test whether the
Ghirlanda/Liang-Zhang relations or the fact of a quasi-universal $\varepsilon$
are more fundamental.

\begin{acknowledgements}
We acknowledge the use of the public data from the {\em Swift} and {\em Fermi}
data and GCN circulars archive. We thank helpful comments from the referee.
This work is supported by the National Natural Science Foundation of China
(Grants 11025313, 10873002, 11078008), the National Basic Research Program
(``973" Program) of China (Grant 2009CB824800), Guangxi SHI-BAI-QIAN project
(Grant 2007201), Guangxi Science Foundation (2010GXNSFC013011), the program for
100 Young and Middle-aged Disciplinary Leaders in Guangxi Higher Education
Institutions, and the research foundation of Guangxi University. It is also
partially supported by NASA NNX09AT66G, NNX10AD48G, and NSF AST-0908362.
\end{acknowledgements}




\begin{deluxetable}{llllllllllllll}

\tablewidth{500pt} 
\tabletypesize{\footnotesize}

\tablecaption{The sample of 29 GRBs for $T_{90,z} <$ 2 s in the rest frame. }

\tablenum{1}
\tablehead{\colhead{GRB}&\colhead{$z$}&\colhead{$T_{90,z}$}&\colhead{$\rm EE$}&
\colhead{$E_{\rm p}$}&\colhead{$E_{\gamma,{\rm iso}}$}&\colhead{log
$\varepsilon$}&\colhead{Type} &\colhead{Afterglow}&\colhead{Refs.}
\\\colhead{name}&\colhead{redshift}&\colhead{(s)}&\colhead{(s)}&
\colhead{(KeV)}&\colhead{($10^{52}$ erg)}&\colhead{}&\colhead{}
&\colhead{X/O(IR)}&\colhead{}}

\startdata
050406 &2.44      &1.57      &n/a      &$25^{+25}_{-19}$    &$0.14^{+0.13}_{-0.03}$         &$-0.74^{+0.73}_{-0.55}$ &high-$\varepsilon$ (II)&Y / Y &(1)\\
050416A&0.6535    &1.51      &n/a      &$15^{+2.5}_{-2.5}$  &$0.1^{+0.01}_{-0.01}$      &${0.00}^{+0.12}_{-0.12}$ &high-$\varepsilon$ (II)&Y / Y &(2)\\
050509B&0.2248    &0.06      &n/a      &$82^{+611}_{-80}$   &$2.4^{+4.4}_{-4}\times10^{-4}$ &$-3.62^{+1.21}_{-0.51}$&low-$\varepsilon$ (I) &N / N &(1,3)\\
050709 &0.1606    &0.09      &130$\pm$7&$83^{+18}_{-12}$    &$2.7^{+1.1}_{-1.1}\times10^{-3}$&$-2.54^{+0.16}_{-0.11}$&low-$\varepsilon$ (I)&N / N&(1,3)\\
050724\tablenotemark{*} &0.2576    &2.39      &154$\pm$1&$110^{+400}_{-45}$  &$9^{+19}_{-2}\times10^{-3}$    &$-2.28^{+0.74}_{-0.30}$&low-$\varepsilon$ (I) &Y / N &(1,3)\\
050813 &$\sim$0.72&$\sim$0.26&n/a &$210^{+710}_{-130}$&$1.5^{+2.5}_{-0.8}\times10^{-2}$ &$-2.75^{+1.12}_{-0.45}$&low-$\varepsilon$ (I)&N / N &(1,3)\\
050922C&2.198     &1.41      &n/a      &$130^{+35}_{-35}$   &$5.3^{+1.7}_{-1.7}$            &$-0.31^{+0.19}_{-0.19}$ &high-$\varepsilon$ (II)&Y / Y &(2)\\
051221A&0.5464    &0.91      &n/a      &$402^{+72}_{-93}$   &$0.28^{+0.21}_{-0.1}$          &$-1.88^{+0.13}_{-0.17}$ &low-$\varepsilon$ (I) &Y / N &(3)\\
060206 &4.048     &1.51      &n/a      &$78^{+9}_{-9}$      &$4.3^{+0.9}_{-0.9}$            &$-0.36^{+0.10}_{-0.10}$ &high-$\varepsilon$ (II)&Y / Y &(2)\\
060502B&0.287     &0.10      &n/a      &$340^{+720}_{-190}$ &$3^{+5}_{-2}\times10^{-3}$     &$-3.59^{+1.14}_{-0.41}$&low-$\varepsilon$ (I) &N / N &(1,3)\\
060614\tablenotemark{*} &0.1254    &$\sim$4.44&106$\pm$3&$302^{+214}_{-85}$  &$0.24^{+0.04}_{-0.04}$         &$-1.68^{+0.51}_{-0.21}$&low-$\varepsilon$ (I) &Y / Y &(2,3)\\
060801 &1.131     &0.23      &n/a      &$620^{+1070}_{-340}$&$0.17^{+0.021}_{-0.021}$      &$-2.64^{+1.01}_{-0.40}$ &low-$\varepsilon$ (I) &Y / N &(1,3)\\
060926 &3.208     &1.90      &n/a      &$19^{+8}_{-18}$     &$0.42^{+0.59}_{-0.04}$    &$0.16^{+0.31}_{-0.69}$  &high-$\varepsilon$ (II)&Y / N &(1,4)\\
061006 &0.4377    &$\sim$0.35&120      &$640^{+144}_{-227}$ &$0.22^{+0.12}_{-0.12}$    &$-2.26^{+0.16}_{-0.26}$  &low-$\varepsilon$ (I) &Y / N &(1,3)\\
061201 &0.111     &0.54      &n/a      &$873^{+458}_{-284}$ &$0.018^{+0.002}_{-0.015}$ &$-3.39^{+0.38}_{-0.24}$  &low-$\varepsilon$ (I) &Y / N &(1,3)\\
061210 &0.4095    &0.57      &85$\pm$5 &$540^{+760}_{-310}$ &$0.09^{+0.16}_{-0.05}$    &$-2.51^{+0.82}_{-0.42}$  &low-$\varepsilon$ (I) &N / N &(1,3)\\
061217 &0.827     &0.22      &n/a      &$400^{+810}_{-210}$ &$0.03^{+0.04}_{-0.02}$    &$-2.96^{+0.47}_{-0.38}$  &low-$\varepsilon$ (I) &N / N &(1,3)\\
070429B&0.9023    &0.25      &n/a      &$120^{+746}_{-66}$  &$0.03^{+0.01}_{-0.01}$    &$-2.13^{+0.75}_{-0.40}$  &low-$\varepsilon$ (I) &N / N &(1,3)\\
070506 &2.31      &1.30      &n/a      &$71^{+95}_{-25}$    &$0.26^{+0.17}_{-0.05}$    &$-1.20^{+0.67}_{-0.26}$  &high-$\varepsilon$ (II)&Y / N  &(1,5)\\
070714B&0.9225  &$\sim$1.56 &$\sim$100 &$1120^{+780}_{-380}$&$1.16^{+0.41}_{-0.22}$    &$-2.16^{+0.51}_{-0.25}$  &low-$\varepsilon$ (I) &Y / Y &(1,3)\\
070724A&0.457     &0.27      &n/a      &$\sim$68            &$0.003^{+0.001}_{-0.001}$ &$-2.51^{+0.24}_{-0.24}$  &low-$\varepsilon$ (I) &Y / N &(3)\\
071020 &2.142     &1.34      &n/a      &$323^{+51}_{-51}$   &$9.5^{+4.3}_{-4.3}$       &$-0.70^{+0.11}_{-0.11}$  &high-$\varepsilon$ (II)&Y / N &(2)\\
071227 &0.394     &1.30      &$\sim$100&$1000^{+200}_{-200}$&$0.22^{+0.08}_{-0.08}$    &$-2.56^{+0.14}_{-0.14}$  &low-$\varepsilon$ (I) &Y / N &(3)\\
080520 &1.545     &1.10      &n/a      &$\sim$30            &$0.07^{+0.02}_{-0.02}$ &$-0.94^{+0.16}_{-0.16}$ &high-$\varepsilon$ (II)&Y / N &(6)\\
080913 &6.7       &1.04      &n/a      &$121^{+232}_{-39}$  &$7^{+1.81}_{-1.81}$       &$-0.77^{+1.39}_{-0.23}$  &high-$\varepsilon$ (II)&Y / Y &(4)\\
090423 &8.1       &1.13      &n/a      &$54^{+22}_{-22}$    &$10^{+3}_{-3}$            &$-0.11^{+0.30}_{-0.30}$  &high-$\varepsilon$ (II)&Y / Y &(7,8)\\
090426 &2.609     &0.33      &n/a      &$45^{+57}_{-43}$    &$0.42^{+0.59}_{-0.04}$    &$-0.72^{+0.92}_{-0.69}$  &high-$\varepsilon$ (II)&Y / Y &(9)\\
090510 &0.903     &0.26      &n/a      &$4414^{+420}_{-420}$&$3.8^{+2.5}_{-2.5}$       &$-2.63^{+0.07}_{-0.07}$  &low-$\varepsilon$ (I) &Y / N &(10)\\
100206 &0.41      &0.12      &n/a &$439^{+73}_{-60}$&$6.15^{+0.39}_{0.39}\times
10^{-2}$ &$-2.61^{+0.07}_{-0.07}$&low-$\varepsilon$ (I) &Y / N &(11)
\enddata \tablenotetext{*}{GRB 050724 and GRB 060614 have $T_{90,z}$ slightly
longer than 2 s. They are included due to the close analogy with other GRBs in
the sample.}

\tablerefs{(1) Butler et al. (2007); (2) Amati et al. (2008); (3) Zhang et
al. (2009); (4) Piranomonte et al. (2006); (5) Thoene et al. (2007); (6) Sakamoto
et al. (2008); (7) Tanvir et al. (2009); (8) Salvaterra et al. (2009); (9)
Levesque et al. (2009); (10) Abdo et al. (2009); (11)Cenko et al. (2010)}
\end{deluxetable}
\newpage


\begin{figure}
   \includegraphics[width=9.0cm,height=9cm]{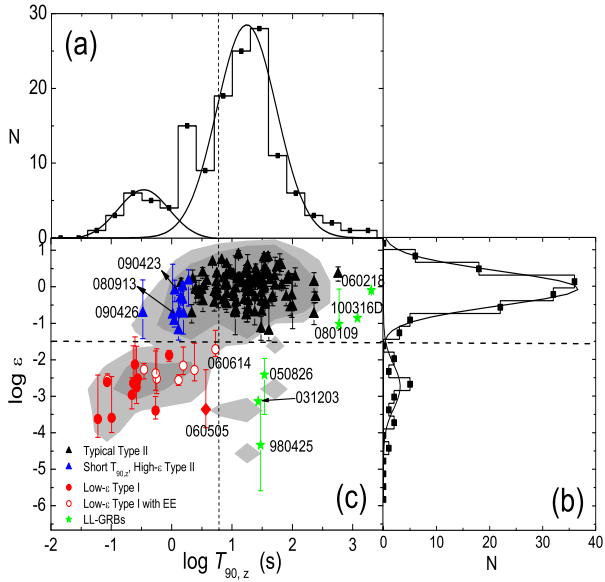}
\caption{The 1-D and 2-D distributions for the bursts in our sample in the
$\log T_{90,z}$ {\em vs} $\log \varepsilon$ plane along with the bimodal fits
({\em solid lines, in panel a, b}) as well as the 2-D distribution for the GRBs
in our sample in the $\log \varepsilon-\log T_{90,z}$ and $\log
\varepsilon-\log (1+z)$ plane ({\em panel c}). The {\em triangles} denote the
Type II GRB candidates as discussed in literature. Among them the traditional
long GRBs are denoted as {\em black} and the GRBs with $T_{90,z}<2$ s are
denoted as {\em blue}. The {\em red solid (open) circles} represent the Type I
GRBs with (without) extended emission as discussed in the literature. The {\em
diamond} denotes the special burst GRB 060505. Nearby LL-GRBs are denoted as
green {\em stars}. The possibility contours for GRB clustering are marked, with
$p\ (\log \varepsilon, \log T_{90,z})> 0.1$ ({\em dark grey}) and $p\ (\log
\varepsilon, \log T_{90,z})> 0.003$ ({\em light grey}), respectively. The
dashed vertical and horizontal lines are the divisions of $\varepsilon=0.03$
and $T_{90,z}= 5$ seconds, respectively.} \label{Fig:fig1}
\end{figure}

\begin{figure}
   \includegraphics[width=9.0cm, angle=0]{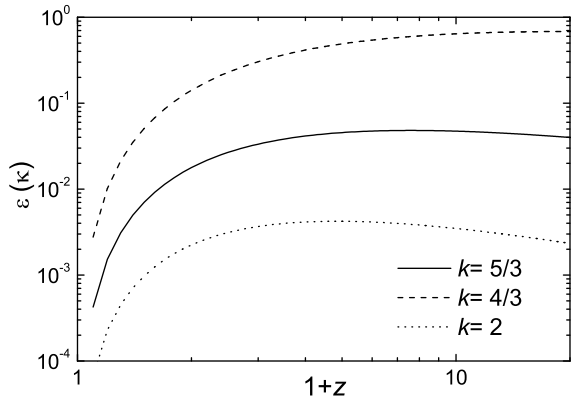}
   \caption{Dependence of $\varepsilon(\kappa)$ on $(1+z)$ for $\kappa=4/3, \ 5/3,\
   2.0$ for a given burst with the same measured fluence and $E_p$.}
   \label{Fig:fig2}
\end{figure}

\begin{figure}
   \includegraphics[width=4.5cm,height=7.cm]{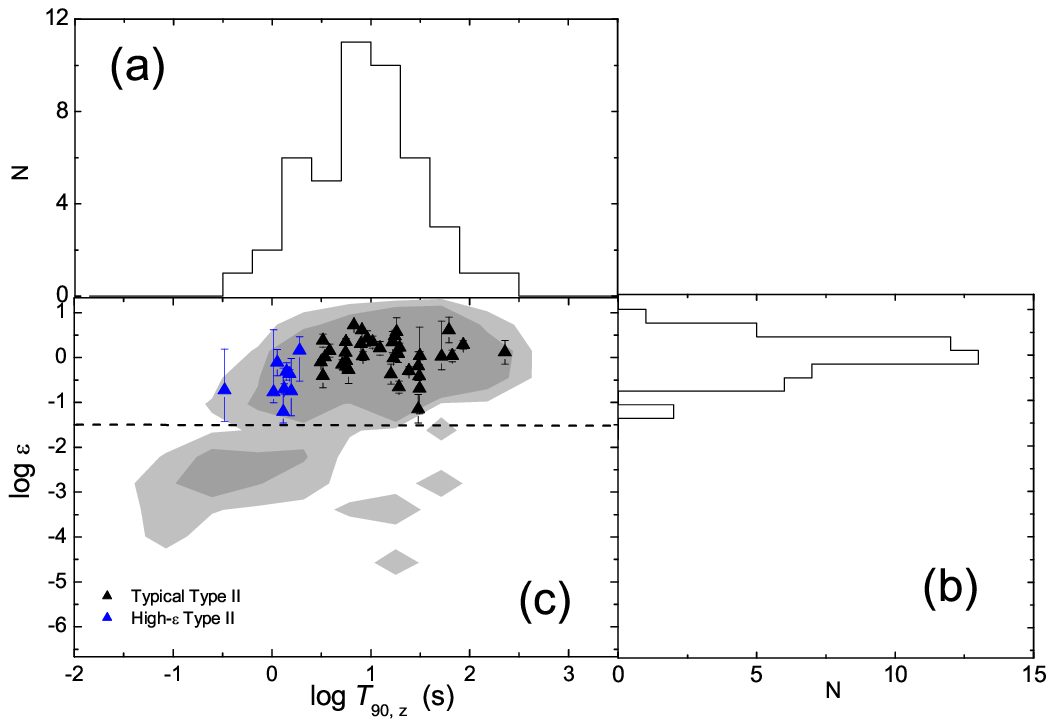}
   \includegraphics[width=4.5cm,height=7.cm]{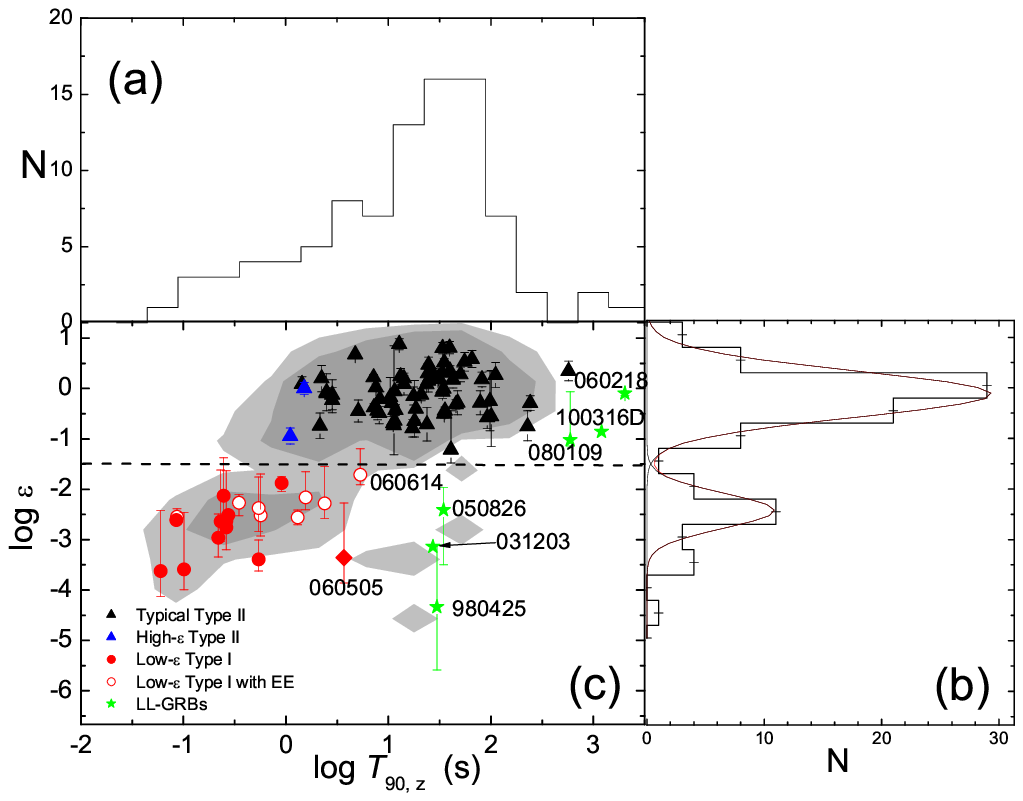}
  \includegraphics[width=4.5cm,height=7.cm]{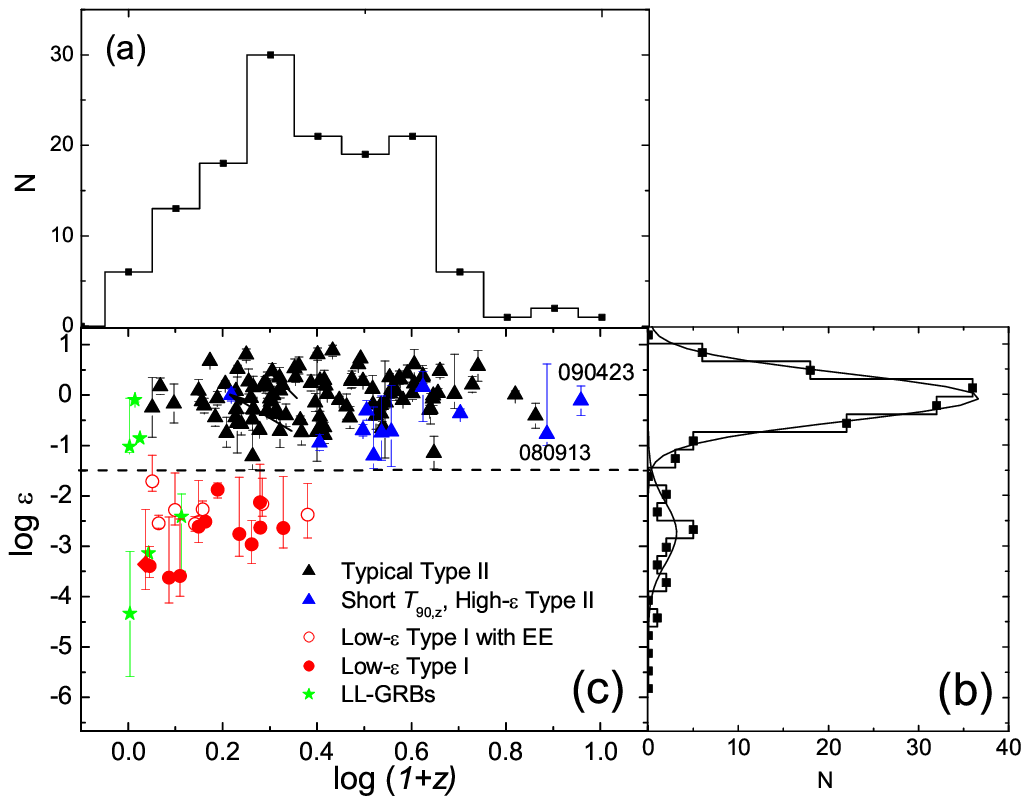}
\caption{The 1-D ({\em panels a and b}) and 2-D ({\em panel c}) distributions
for the bursts in our sample in the $\log T_{90,z}-\log \varepsilon$
 plane for GRBs for $z>2$ ({\em
left group of panels}) and $z<2$ ({\em middle group of panels}). The GRB
distributions in the $\log (1+z)-\log \varepsilon$ plane are also shown in the
{\em right group of panels}. The symbol style is the same as Figure 1. The
contours are the same as that in Figure 1 for all GRBs in our sample.}
  \label{Fig:fig4}
\end{figure}

\begin{figure}
   \includegraphics[width=7.0cm,height=7cm]{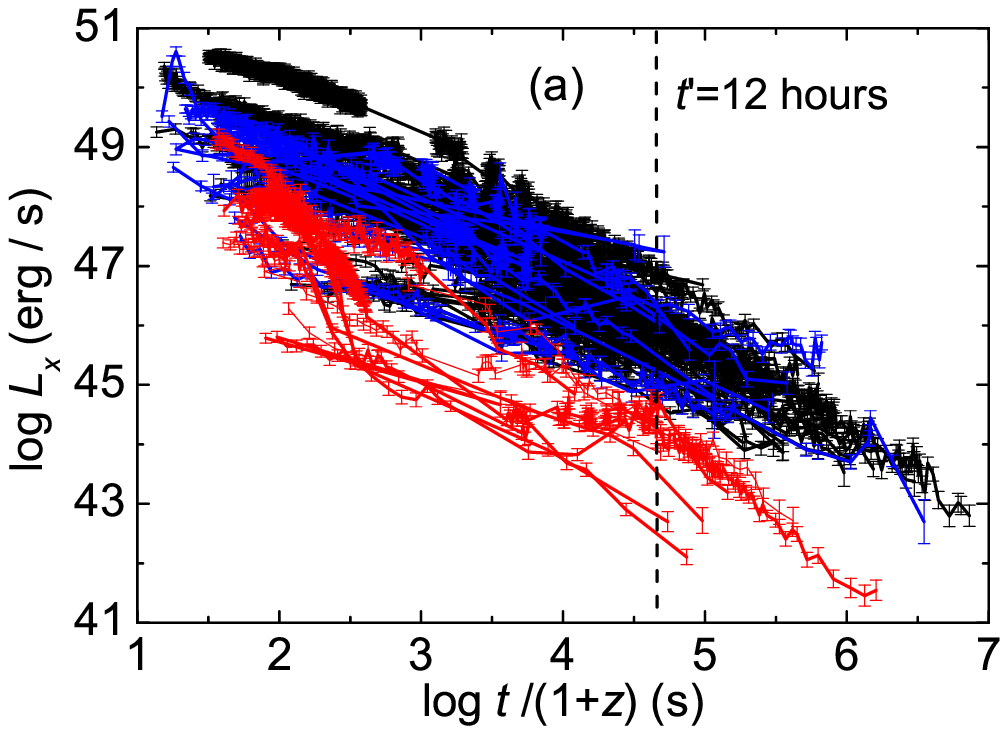}
   \includegraphics[width=7.0cm,height=7cm]{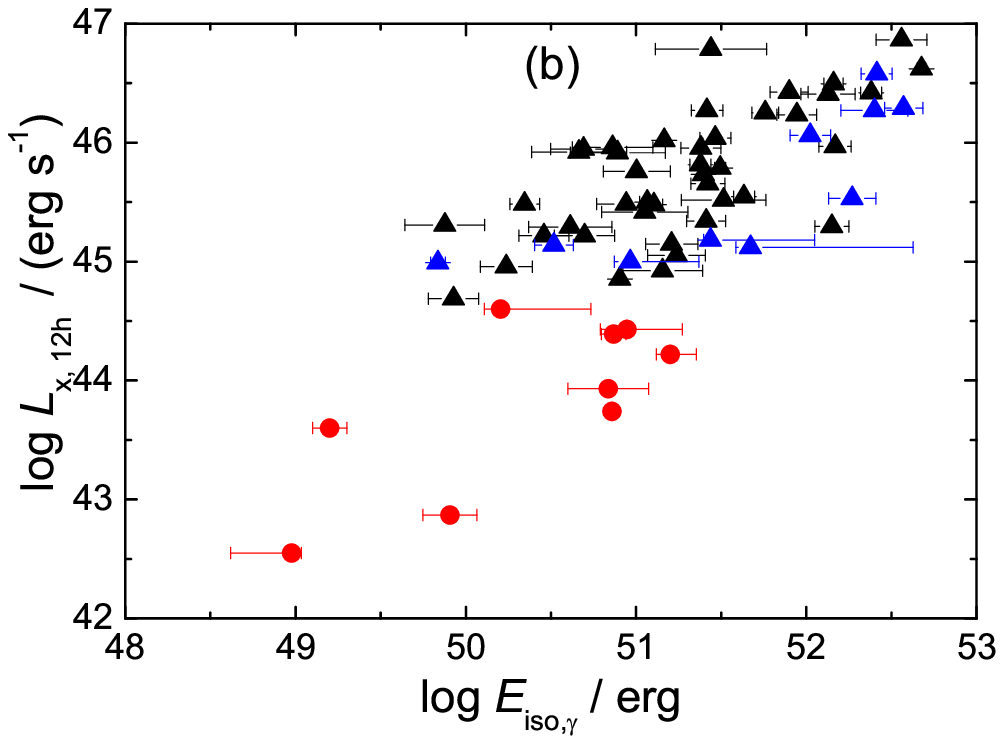}
  \caption{Comparisons of (a) the XRT lightcurves and (b) the $L_{X,12h}-E_{\rm iso}$
distribution among the low-$\varepsilon$ GRBs with $T_{90,z}<5$ s ({\em red}),
high-$\varepsilon$ GRBs with $T_{90,z}<2$ s ({\em blue}), and 44 long-duration
high-$\varepsilon$ GRBs from the Liang et al. (2009) sample ({\em black}). Here
$L_{X,12h}$ is the X-ray luminosity in the XRT band at the rest-frame 12 hours
post the GRB trigger.}
  \label{Fig:fig4}
\end{figure}

\end{document}